# Offensive tool determination strategy
## R.I.D.D.L.E. + (C)

Herman Errico [1]

[1] Italian Association of Critical Infrastructures' Experts, Rome (RM), Italy

**Abstract.** Intentional threats are a major risk factor related to infrastructures critical asset vulnerabilities, therefore an accurate risk assessment is necessary to analyze threats, assess vulnerabilities and evaluate the impact on assets, infrastructures or systems. For this reason, the present research aims at describing a specific methodology that can be included as an additional phase in the risk assessment process. The introduction of an additional analysis' parameter concerning offensive tools characteristics is proposed to foster a better understanding about intentional threats. The description of such methodology is based on a simple and comprehensible language that could be used by a broad and non-sectorial audience. This methodology is set up on an approach described as "offensive tool determination strategy", with the acronym R.I.D.D.L.E. + (C) which indicates the variables' names (resistance, intrusion timing, damage, disruption timing, latency, efficiency and cost) used as parameter to perform the analysis, supported by open source intelligence. As a result, a specific range of values, assumed from their impact on the targeted asset, are assigned to every variable. Subsequently, a matrix is provided for practical application, which could leak out unexpected vulnerabilities on assets' security and provide a more granular framework for decision-making support.

## 1 Introduction

The factor of risk definition we intend to analyse is threat, which could be defined as "potential cause of an unwanted incident, which may result in harm to a system or organization" [1] or as "any circumstance or event with the potential to adversely impact an asset through unauthorized access, destruction, disclosure, modification of data, and/or denial of service" [2].

Threats to an infrastructure critical asset can originate from different threat sources. This research focuses its attention on human intentional threat, as it represents the most frequent and dangerous threat. The aforementioned statement is confirmed by "ENISA Threat Taxonomy" [3], in which "Physical attack (deliberate/intentional)", defined as "threats of intentional, hostile human actions", are presented in first place on the "high level threats" [3], and by "Worldwide Threat Assessment of the US intelligence Community" [4] where cyber threat and Terrorism are a major threat for National Secuirty. Data exfiltration, vulnerabilities exploitation, service disruption, theft of company property and many other consequences of hostile actor's behaviours should be prioritized during an integrated risk management process [5], particularly in the risk assessment [6] phase, in order to quantify and identify risk aspects related to intentional threats.

Concentrating on the aggressor's capability and ability to attack a specific target could help anticipating and detecting unexpected vulnerability on assets' security, in a system or an infrastructure, that a security manager [7] or other individuals involved in a risk assessment process of risk identification, risk analysis and risk evaluation didn't notice. Such line of reasoning brought this research to consider offensive tools as the key factor of intentional threat analysis, defining an offensive tool determination strategy. Specifically, this method aims at proposing an additional component, suitable for multiple risk assessment methodology, which is supported by open source intelligence analysis, and can be applied to critical infrastructure or any other organization.

This methodology is denominated "offensive tool determination strategy", identified with the acronym of "R.I.D.D.L.E. + (C)", which indicates the names of the variables used in the analysis: resistance, intrusion timing, damage, disruption timing, latency, efficiency and cost. Those variables are selected as a result of the observation of known cyber and kinetic tools through a research based on information processing matured from computer science, and supported by open source intelligence, which led to the analysis of their respective characteristics. Subsequently, we assigned a specific range of values to every variable that describes its impact on the target, then we constructed a matrix for practical application. This method should be able to simulate the tools' selection operated by a hostile actor, to prevent or at least mitigate the effects of a potential security breach or threat.

## 2 State of the art of risk assessment techniques

In scientific literature, different risk assessment techniques can be observed, which are based on different approaches. Most of these techniques are based on risk identification, risk analysis and risk evaluation, specifically directed to the description and classification of threats, detection of vulnerabilities and evaluation of impact. Our concern is on risk analysis and more specifically on the understanding of offensive tools used as attack vectors in intentional human threats. Various risk assessment techniques will be described, some of which do not show strong attention to offensive tools in risk analysis, but only to threats in general, to other techniques that deepen the issue of threats, also focusing on offensive tools. Conducting a risk assessment by taking into account these elements will be crucial for developing a risk mitigation strategy based on a more granular framework for decision-making.

The ISO 27000 Family Standards considers various risk assessment techniques, with ISO / IEC 27001:2013 focused mainly on risk-based planning [8]. Since the 2013 revision does not require so-called asset-based risk assessment, which would identify the risks based on assets, threats and vulnerabilities – according to ISO27001:2013, your company can identify risks using some other (less complicated) method. This is to ensure that the identified information risks are appropriately managed according to threats and their nature. You can identify risks based on your processes, your departments, focusing on threats instead of vulnerabilities, or any other methodology you like. The 2013 revision has made risk management for information Secuirty easier to read and understand, and ensures compliance with the principles of risk management contained in ISO/IEC 31000:2009 [9]. For what concerns ISO/IEC 31000:2009 the risk assessment technique, provided by ISO/IEC 31010:2009 [10], which support threat analysis are "Scenario Analysis" (Annex B.10) [10] and the Consequence/probability matrix (Annex B.29) [10], but none of them analyses offensive tool. While assets-threats-vulnerabilities methodology provided by ISO/IEC 27001:2005 [11], and specifically in ISO/IEC TR 13335-3 at 9.3.4 point (Assessing threats) [12], is no longer required under ISO 27001:2013, it still represents a robust methodology in order to conduct risk assessment. you need to list all your assets, then threats and vulnerabilities related to those assets, assess the impact and likelihood for each combination of assets/threats/vulnerabilities and finally calculate the level of risk.

The ITU-T X.1205 recommendation "Overview of Cybersecuirty" [13] provides a much more threat oriented risk analysis. This statement is confirmed at point 7.2 where "any element of the cyber environment can be viewed as a security risk, which is generally thought of as a combined assessment of threat. Threat analysis includes the task of describing the type of possible attacks, potential attackers and their methods of attack and the consequences of successful attacks. […] "Risk assessment combined with threat analysis allows an organization to evaluate potential risk to their network" [13], and in Appendix I where Attackers techniques are analysed, but in a descriptive way as it "does not form an integral part of this Recommendation" [13].

The National Institute of Standards and technology has released two special publications on information security management. "Security and privacy controls for federal Information systems and Organization" through special publication 800-53 r4 [14] and "Managing Information Security Risk Organization, Mission, and Information System View" through special publication 800-39 [15], the first is oriented to the "selection and implementation of *security controls* for information systems and organizations" [14] and the other "is to provide guidance for an integrated, organization-wide program for managing information security risk to organizational operations (i.e., mission, functions, image, and reputation), organizational assets, individuals, other organizations, and the Nation resulting from the operation and use of federal information systems" [15]. Both special publications refer to NIST Special Publication 800-30 r1 "guide for conducting risk assessments" [16] which provides a specific and detailed process for conducting a risk assessment (Step 2) beginning by the identification of the threat source (Task 2-1) and threat events (Task 2-2). This risk assessment technique could be considered as mostly threat oriented, despite the lack of offensive tools characteristics analysis, there is a specific description of input for threat events in table E-2 [16].

In the NATO Cooperative Cyber Defence Centre of Excellence "National Cyber Security Framework Manual" [17] there is a specific mention of risk assessment although it is referred to National Secuirty State level by applying a National Risk Assessment (NRA) in order to "reduce the structural causes of insecurity" [17] and to maintain a National Cyber Security strategy for the "alignment with the critical (information) infrastructure protection strategy (C(I)IP)," [17]. This political framework provides the asset-threat-vulnerability techniques provided by ISO/IEC 27001:2005 despite focusing on the evaluation of the various 'attacks' (which includes intentional and unintentional acts of human and natural origin) that a system can be subjected to.

Regarding Critical Infrastructure Protection (CIP) "there is a significant number of risk assessment methodologies for critical infrastructures. In general the approach that is used is rather common and linear, consisting of some main elements: Identification and classification of threats, identification of vulnerabilities and evaluation of impact."[18]

Concerning Europe, in 2006 the European Commission provided the "European Programme for Critical Infrastructure Protection" (EPCIP) [19] followed by the release of Council Directive 2008/114/CE [20] and the implementation of the Critical Infrastructure Warning Information Network (CIWIN) [21]. In 2013 the European Commission published a document called "new approach to the European Programme for Critical Infrastructure Protection Making European Critical Infrastructures more secure" [22], this document determines a revised implementation of the EPCIP and provides a new approach based on the "practical implementation of activities under the prevention, preparedness and response work streams." [22] After identifying four critical infrastructures, such as Eurocontrol, Galileo, the electricity transmission grid and the gas transmission network, the aim of this new approach is to work with those four CI's to "set up tools for risk assessment and risk management, taking stock of existing research and innovation activities"[22].

Following this path, in 2015, the European Joint Research Center, published the "Risk assessment methodologies for critical infrastructure protection. Part II: A new approach" [24], based on the aforementioned document and the paper of 2010 "Risk Assessment and Mapping Guidelines for Disaster Management" [23] from the European Commission, based on the framework of ISO/IEC 31000 approach and more attentive to natural disasters. The risk assessment methodology proposed by JRC in 2015 aims at providing a "Critical Infrastructures & Systems Risk and Resilience Assessment Methodology" (CRISRRAM) [24] which is based on a system of system approach at various levels (asset level, system level and society level) and considers an "all-hazard approach" (natural, human and accidental), by underlining the necessity of tools that focus "on designing scenarios" [24]. This methodology doesn't seem to address particular attention to offensive tools but it provides a description of threat likelihood impact.

As a conclusion, we can say that in the state of the art of risk assessment techniques there is a common practice of conducting risk analysis, which is based on the threat description and classification, but there is a gap in the offensive tools analysis utilized by intentional human threat for perpetrating an attack. For this reason, it is our intention to try to provide a simple analytical tool for implementing a more granular risk analysis and try to fill this gap.

## 3    Offensive tools' analysis

In order to provide a concrete basis to choose this type of process, it will be necessary to focus on the research-based problem: security. The deep knowledge of offensive tools, which can produce negative direct effects on our infrastructure assets, is of utmost important for the protection of our assets. Specifically, in support of our research, we will conduct an open source intelligence analysis on the most common offensive tools, and we will analyze them applying engineering processes, at a conceptual level, exploiting their observation methods.

To be able to extract the variables for conceiving the matrix of "R.I.D.D.L.E. + (C)" we utilized a strategy based on information processing by applying reverse engineering [25] process on a selection of cyber and kinetic tools.

Using reverse engineering to observe the characteristics and vulnerabilities of offensive tools is due to its use for security purpose, specifically in ICT security.

The method we used derives from the combination of two elements of reverse engineering: the system level and the code-level. In this way, both general and specific aspects of each tool can be observed.

In summary, we wanted to use and apply a top-down strategy by the static observation of their common characteristics.

Therefore, we borrowed a technical tool from reverse engineering, analyzed his functioning and created an analysis tool. The reverse engineering tool is a disassembler [26], which specifically modifies a binary executable program and transforms it into a text file that contains the code that built the program. The implementation of those procedure led our vision to create a general personalized technique, which could help us to determine tools' characteristics considering all their types, working principles and vulnerabilities, named "tools observation table".

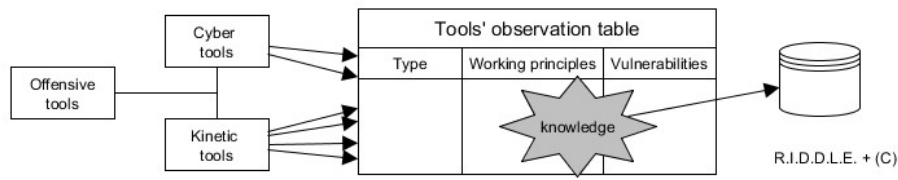

The "tools observation table" is an essential method where to insert data deducted from open source intelligence analysis, same process applied by hostile actors to conceive targeting operations to carry out their attacks.

The logical path that led us to extract the variables involved in "R.I.D.D.L.E. + (C)" originated from a first OSINT analysis of cyber security [27-30] and physical security [31,32] threats. The result of the aforementioned analysis led our research to identify the most common offensive tools characteristics, afterwards we gathered them into their general category. Regarding the cyber offensive tools, we considered: virus, worm, trojan horse, remote access tools, malicious code, for whom we analysed their different classes, working principles and vulnerabilities. For what concerns the kinetic offensive tools, we took into account those which could cause destruction or misappropriation of company properties or a denial of service: explosive attacks, vandalism, chemical attack, perimeter breach, diversion, destruction or sabotage of supply structure and armed assault. As a final result of this analysis stage, we were able to extract the R.I.D.D.L.E. + (C) variables.

## 4   Methodology description

"R.I.D.D.L.E. + (C)" is an offensive tool determination strategy for risk assessment process that can be utilized by critical infrastructure or any other infrastructure that needs to secure its assets, so that it can provide useful guidance concerning the type of attack that most likely might be conducted towards an infrastructure critical asset.

This allows us to provide for technical defense measures or to observe the vulnerabilities that would not be previously identified.

This method allows operators to simulate the attacker behaviours during the phase of available tools' selection. Thinking like our adversary could give us a different perspective of risk assessment process, specifying threats by considering known tools or hypothesizing possible tools. Applying logical analysis allows us to reveal the kind of patterns or correlations that a technological tool might miss.

In addition, it would allow non-technical audience to understand, for simplicity of language, the world of risk analysis, but also the understanding of the cyber security sector, specifically related to the cyber threat.

### 4.1   Variables description and ranking

Considering the selected variables, we became aware of the necessity to divide the variables in multiple stages of values, in order to represent the different strength or weakness of the tools' characteristics. We used an ordinal level of measurement in which we represented five intervals (from 1 to 10) related to the description of the variable intensity.

- Resistance: The ability of an offensive tool to withstand all the attempts taken and the forces applied by the target to obstruct, block or prevent the attack [31]. It represents the measure of its strength and its vulnerabilities, depending on the ability of the attacker to move forward in the perpetration of the attack despite the difficulties encountered in attempting to achieve its ultimate goal [32]. Such resistance is related to the tool's intrinsic power, accuracy, sophistication and adaptability, to the predictability of its use, and the ability of the attacker in the accurate choice of the proportional and more suited one to the predetermined goal [33].

| Resistance variable | Score |
|---|---|
| Extreme resistance. The offensive tool can withstand attempts to stop or destroy it. It is almost impossible to interrupt its action or, even if | 9-10 |
| blocked, due to its heavy resistance it can still produce the effects it was conceived for. | |
| High resistance. The offensive tool resists attempt to stop or destroy it, but the produced effects are smaller than when the attack started. | 8-7 |
| Medium resilience. The instrument has an average resistance and is able to accomplish the attack; its effects are mediocre, but still of concern. | 6-5 |
| Low resistance. The offensive tool has a low resistance, after several attempts to block it or destroy it, its action is arrested and it can't accomplish the action for which it was conceived or it can be of low relevance. | 4-3 |
| None resistance. The offensive tool is not resistance and, since the first attempt to stop it or destroy it, does not retain the capabilities it had been designed for. | 2-1 |

- Intrusion timing: Intrusion timing represents the time measurement that the tool employs and needs to finalize the attack and reach the target. It depends on the strength of the tool and the robustness of the target security system designed to prevent and neutralize any potential risk [34]. If the timing intrusion is short, the instrument immediately accesses and increases the chance of success for the attacker. If, on the other hand, the intrusion timing is long, the tool could take a long time to access, proportionally increasing for the attacker the risk of failure [35].

| Intrusion timing variable | Score |
|---|---|
| Immediate intrusion time. The instrument can access the System or the Infrastructure instantaneously, the time segment is between 1 second and 10 seconds. | 9-10 |
| Short intrusion time. The instrument is able to enter the System or Infrastructure after a short time, the time segment is within 20 seconds and 1 minute. | 8-7 |
| Medium intrusion time. The instrument can enter the System or Infrastructure over an average length of time, the time range is between 1 and 12 hours. | 6-5 |
| Long intrusion time. The instrument manages to penetrate the System after a long period, the time range is between 1 day and 1 week. | 4-3 |
| Very long intrusion time. The instrument manages to penetrate the System after a very long period, the intrusion time range is longer than one week. | 2-1 |

- Damage: Damages represent the short, medium, and long-term impact on the asset vulnerability consequences of the attack on the target. The different types are related to economic, reputational and psychological nature. The damage may concern material and physical goods (such as real estate, machinery, territorial areas) or intangible assets (such as: monetary / financial, image or reputation of an enterprise or a person, future business, business, profitability) [36]. Damage can therefore be related to the quality, adequacy, security, availability of the service or goods delivered by the target attack and the resulting image and integrity of the target, but they may also reflect on the population and the social order (in terms of victims, suffering Moral and physical, sectoral, public and national security) [31]. They depend on the type and range of action and the intrinsic force of the tool used to perpetrate the attack, but also on the resulting aggressiveness and pervasiveness of the attack and the relevance and resilience of the chosen target. In a context of high interdependence between infrastructures, a failure caused by human action consisting in a physical or cyber-attack against a critical infrastructure can easily produce domino or cascade effects and rapidly extend to other critical infrastructures, amplifying total damage and malfunctioning up to causing a catastrophic crisis of the entire national system [37].

| Damage variable | Score |
| --- | --- |
| Severe damage. The tool can produce very serious damages. The resulting effects affect the physical infrastructure, the network or the organization to the extent of 90 to 100%. Such damages are either irreparable or difficult to remedy. | 9-10 |
| High damage. The tool can produce serious damage. The resulting effects affect the physical infrastructure, the network or the organization to the extent of 70 to 90%. Such damages are repairable in the long-term. | 8-7 |
| Medium damage. The tool can produce contained damage. The resulting effects affect the physical infrastructure, the network or the organization to the extent of 40 to 70 %. Such damages are repairable in the medium-term. | 6-5 |
| Low damage. The tool can produce lowered damage. The resulting effects affect the physical infrastructure, the network or the organization to the extent of 10 to 40 %. Such damages are repairable in the short-term. | 4-3 |
| Minimum damage. The tool can produce damage. The resulting effects affect the physical infrastructure, the network or the organization to the extent of 1 to 10 %. These damages can be immediately repaired. | 2-1 |

• Disruption timing: It measures the duration of the suspension or cease of the service caused by the offensive tool, which causes problems with its availability and its functionality. This interruption consists in the discontinuity of actions, processes, related to target activity, caused by the effective use of the instrument chosen to attack [38].

| Disruption timing variable (*) | Score |
|---|---|
| Very long disruption. The effects of the instrument's action consist in interrupting the service for more than six months. | 9-10 |
| Long disruption. The effects of the instrument's action consist in interrupting the service for a time period between six and three months. | 8-7 |
| Medium disruption. The effects of the instrument's action consist in interrupting the service for a time period between three months and one week. | 6-5 |
| Short disruption. The effects of the instrument's action consist in interrupting the service for a time period ≤ than one week. | 4-3 |
| Minimum disruption. The effects of the instrument's action consist in interrupting the service for a time period ≤ than one day. | 2-1 |

*(Descriptions of disruption measures should be altered in the presence of a cyber tool, extending the time segment from one hour to a week).

• Latency: Latency means the time segment between the time of intrusion of an offensive tool within the system and the time when that tool is identified by the system security elements and the propagation or duplication of the effects of its use within the system [33]. Specifically, this feature expresses the stealth mode of the instrument, which represents a series of technical or camouflage arrangements that make the instrument imperceptible, unidentifiable, not intercepted, as well as intrusion, even in the moment of permanence and carrying out the action within the structure [38]. The greater the latency, the greater the value to be attributed to this measure, since it can act undisturbed within the structure and produce the malicious effects for which this mode has been developed.

| Latency variable | Score |
|---|---|
| Very Long latency. The instrument is hardly identifiable or intercepted and therefore it is difficult to quantify the latency period and propagation of its effects. | 9-10 |
| Long latency. The tool is able to endure in the system and multiply its effects for a long time before revealing itself. | 8-7 |
| Medium latency. The tool is able to endure in the system and multiply its effects for a discrete time before revealing itself. | 6-5 |
| Short latency. The tool is able to endure in the system and multiply its effects for a short time before revealing itself. | 4-3 |
| Very Short latency. The tool is able to endure in the system and multiply its effects for a very short time before revealing itself. | 2-1 |

• Efficiency: The efficiency of an instrument represents its ability to perform the actions it was designed and conceived for, the greater the ability to succeed, the greater its effects. Specifically, the elements that make up the efficiency are given by the ability to exploit the potentialities that have been attributed to it and the way these potentials come out without waste or loss of resources in the actual service supply [39]. The efficiency of an attack tool represents a high-risk parameter considering that if an instrument is efficient it can produce its effects and then complete the attack in 100% of the cases. We can then classify and evaluate an instrument according to its efficiency rate, which is described in the table below.

| Efficiency variable | Score |
|---|---|
| Highly efficient. The instrument is highly efficient and can produce the effect it has been conceived for in 100% of cases. | 9-10 |
| Discreetly efficient. The instrument is highly efficient and can produce the effect it has been conceived for in 80% of cases. | 8-7 |
| Mediumly efficient. The instrument is highly efficient and can produce the effect it has been conceived for in 60% of cases. | 6-5 |
| Lowly efficient. The instrument is highly efficient and can produce the effect it has been conceived for in 30% of cases. | 4-3 |
| Not efficient. The instrument is highly efficient and can produce the effect it has been conceived for in 10% of cases. | 2-1 |

- Cost: It represents the expense you need to sustain to produce or buy a certain tool. It may depend on its sophistication and accuracy, its technical components, its availability and the know-how needed to originate it or to rip it. The specific cost is parameterized on a reversed value scale as a high-cost tool can be a barrier to a hostile actor, but a low-cost tool is an incentive to acquire that type of tool [40].

| Cost variable | Score |
|---|---|
| Minimum cost. Costs for the purchase or production of the instrument are less than or equal to € 1000. | 9-10 |
| Low cost. Costs for the purchase or production of the instrument are equal to or greater than € 1000. | 8-7 |
| Medium cost. Costs for the purchase or production of the instrument are equal to or greater than € 10000. | 6-5 |
| High-cost. Costs for the purchase or production of the instrument are equal to or greater than € 100000. | 4-3 |
| Very high-cost. Costs for the purchase or production of the instrument are equal to or greater than € 1000000. | 2-1 |

## 5    "R.I.D.D.L.E. + (C)" employment

Before utilizing any appendix and giving any score related to a specific tool, it is imperative that the operators determine the asset that needs to be protected by asking themselves those simple questions: Which asset do you want to secure? From which threats you want to secure them? How many losses you will face if you don't succeed? How many resources are you ready to invest for prevention?

Once you have answered all the previous questions you will face a variety of assets you want to protect or maybe should protect and you will be able to determine what budget would be necessary for risk mitigation.

Subsequently, you will need to establish a team that can perform a OSINT analysis to find out about the possible offensive tools that could be used against our infrastructure and list them. We suggest to insert the results of the OSINT analysis in the "tools' Observation Table", to make the tools' contents emerge.

The produced results will be a useful parameter to score the tools by applying the suggested evaluations to the "R.I.D.D.L.E. + (C)" matrix considered in Appendix 1, then we suggest to write a detailed description of the reasons behind the scoring by using Appendix 2. At the very end of this process you could appreciate the results of the scoring related to every tool and determine if the tool represents either a minor, a medium or severe threat described in Appendix 3.

## 6   Concluding remarks.

At the end of the present work, we can observe that the presence of intentional threat is a major element within the risk assessment. To define the type of intentional threat that could negatively affect our critical assets, we focused on studying the offensive tools, using a reverse engineering technique to analyze them. The method we used to define the problem is based on a static observation of the features of the individual elements, which, together with the OSINT analysis, is a valid tool for us to try to bring out those vulnerabilities that could not be considered as having a Risk assessment analysis focused solely on potential threats to our assets. As a last comment, it is necessary to consider that the research is carried out as the first phase of a project that will be subject to detailed implementation and study, and can aspire to build a complete model of integrated risk management.

Appendix 1

| "R.I.D.D.L.E. + (C)" tool comparison table | | | | | | | | | |
|---|---|---|---|---|---|---|---|---|---|
| Tool name | R | I | D | D | L | E | C | Score | Total |
| | | | | | | | | | |
| | | | | | | | | | |
| | | | | | | | | | |
| | | | | | | | | | |

Appendix 2

| Tool description table | |
|---|---|
| Offensive tool name | |

12| Variable | Score | Motivation | Notes |
|---|---|---|---|
| Resilience | | | |
| Intrusion timing | | | |
| Damage | | | |
| Disruption timing | | | |
| Latency | | | |
| Efficiency | | | |
| Cost | | | |

Appendix 3

| Threat level | Description | Scores |
|---|---|---|
| Severe threat | The severe threat level represents a tool that is capable of causing serious or irreversible damage to infrastructures, so in presence of such high score, it is mandatory to pay close attention and to put in place all the necessary activities to reduce the risk related to the tool potential activity. | 50-70 |
| Medium threat | The medium threat level represents a tool that is capable of causing medium and repairable damages to infrastructures, so in presence of such score, it is necessary to pay a discrete prudence but it is still necessary to reduce the risk related to the tool potential activity. | 25-50 |
| Minor threat | The minor threat level represents a tool that is capable of causing minor or soft and easily repairable damages to infrastructures, so in presence of such low score, it is still necessary to put in place prevention activity to decrease the risk related to the tool potential activity. | 0-25 |

**References**

1. International Organization for Standardization: ISO/IEC 27000:2016 – Information technology – Security techniques – Information security management systems – Overview and vocabulary, ISO, Geneva, Switzerland (2016).